\documentclass[superscriptaddress,reprint,prb,floatfix]{revtex4-2}
\usepackage{graphicx,color,dcolumn,bm}
\usepackage{amsmath}
\usepackage[colorlinks=true,breaklinks=true,allcolors=blue]{hyperref}

\begin{document}
\title{Hole Mobility Calculation for Monolayer Molybdenum Tungsten Alloy Disulfide}
\author{Ming-Ting Wu}
\author{Cheng-Hsien Yang}
\author{Yun-Fang Chung}
\affiliation{Department of Electrical Engineering, National Chung Hsing University, Taichung 40227, Taiwan, R.O.C.}
\author{Kuan-Ting Chen}
\affiliation{Institute of Electro-Optical Science and Technology, National Taiwan Normal University, Taipei 11677, Taiwan, R.O.C.}
\author{Shu-Tong Chang}\email{stchang@dragon.nchu.edu.tw}
\affiliation{Department of Electrical Engineering, National Chung Hsing University, Taichung 40227, Taiwan, R.O.C.}
\date{\today}

\begin{abstract}
A simple band model using higher order non-parabolic effect was adopted for single layer molybdenum tungsten alloy disulfide (i.e., $\mathrm{Mo}_{1-x}\mathrm{W}_x\mathrm{S}_2$). The first-principles method considering $2\times2$ supercell was used to study band structure of single layer alloy $\mathrm{Mo}_{1-x}\mathrm{W}_x\mathrm{S}_2$ and a simple band (i.e., effective mass approximation model, EMA) model with higher order non-parabolic effect was used to fit the first-principle band structures in order to calculate corresponding the hole mobility. In addition, we investigate the alloy scattering effect on the hole mobility of $\mathrm{Mo}_{1-x}\mathrm{W}_x\mathrm{S}_2$.
\end{abstract}

\maketitle

\section{Introduction}\label{sec:1}
Among the typical transition-metal dichalcogenide (TMD) alloys, single layer molybdenum tungsten alloy disulfide has a characteristic atomically thin two-dimensional layered structure that contributes to its unique electronic properties, allowing its application in transistor device technology. As the band gap of TMD alloys can be altered, these alloys can be suitably applied in band-gap engineering for transistor devices. In order to develop novel $p$-type field effect transistor (PFET), it is crucial to consider how the type of TMD alloy used affects its band structure and the corresponding hole mobility. Two types of TMD alloy single layers are discussed in our previous work. As an example of the first type, molybdenum tungsten alloy disulfide (Mo$_{1-x}$W$_x$S$_2$), with its two-dimensional layered structure, has great potential application in future transistor technology. TMD alloys may have better intermiscibility, and the synthesis of TMD bulk alloys with multiple layers, such as Mo$_{1-x}$W$_x$S$_2$~\cite{a01,a02}, imparts good thermodynamic stability to the material. In a previous study~\cite{a03}, the analysis of single layer Mo$_{1-x}$W$_x$S$_2$ by scanning transmission electron microscopy (TEM) revealed its homo-atomic and hetero-atomic coordinates. Moreover, the quantified degree of alloying of the transition metals such as W and Mo was successfully obtained and their alloying mechanism in the single layer was identified. In another recent study~\cite{a04}, the first family of atomic-single layer Mo$_{1-x}$W$_x$S$_2$ was exfoliated and the x-dependent photoluminescence (PL) was observed from the single layer alloy Mo$_{1-x}$W$_x$S$_2$. Further, the composition-dependent physical properties of single layer Mo$_{1-x}$W$_x$S$_2$, including its band structure and effective mass, were determined by the first-principles calculation using a large $9\times9$ supercell with an effective band structure (EBS) approximation model~\cite{a05,a06}.

It was found that as the W composition increases, the effective mass of the hole decreases; this is because of the similar contributions of MoS$_2$ and WS$_2$ to valence bands. Smaller effective mass is favorable for hole mobility. Thus, we aim to explore the hole mobility of the monolayer Mo$_{1-x}$W$_x$S$_2$ for PFET application. A simple band model cannot effectively calculate the valence band structure of single layer alloy Mo$_{1-x}$W$_x$S$_2$. Therefore, development of a compact valence band model for single layer alloy Mo$_{1-x}$W$_x$S$_2$ considering the calculation of hole mobility is crucial. In this study, we propose a compact valence band structure and hole mobility calculation model for the Mo$_{1-x}$W$_x$S$_2$ single layer. Section~\ref{sec:2} presents the $2\times2$ supercell first principle band structure of Mo$_{1-x}$W$_x$S$_2$ single layer, a simple band model, and hole mobility calculation
of single layer Mo$_{1-x}$W$_x$S$_2$. Section~\ref{sec:3} discusses the results of the experiments conducted in this study. Finally, Section~\ref{sec:4} presents the conclusions of this study.

\section{Computation Methods}\label{sec:2}
\begin{figure}[b]
\includegraphics[width=2in]{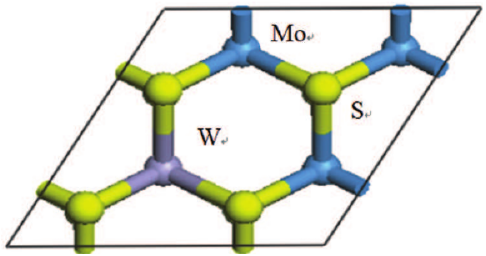}
\caption{Structure of a $2\times2$ supercell of Mo$_{0.75}$W$_{0.25}$S$_2$ (including one W atom and three Mo atoms) in the $x$-$y$ plane. The yellow spheres represent S atoms, and the blue and purple spheres represent Mo and W atoms, respectively.}\label{fig:1}
\end{figure}

\subsection{First-Principles Method}\label{sec:2-1}
In this work, we used the Vienna \textit{Ab Initio} Simulation Package (VASP) considering the projector augmentedwave (PAW) potentials to calculate the band structure of monolayer Mo$_{1-x}$W$_x$S$_2$. Note that density functional theory (DFT) was used with local density approximation (LDA) exchange–correlation functional, and $400$~eV cutoff energy for plane-wave expansion. We also used $21\times21\times1$ k-grid mesh of the supercell, and set the electronic self-consistent cycles of energy convergence criterion to $10^{-5}$~eV. The spin orbit coupling was not including in our work because prediction with the inclusion of spin orbit coupling provided smaller band gap of WS$_2$ than that of MoS$_2$. To model single layer Mo$_{1-x}$W$_x$S$_2$, a $2\times2$ supercell (including $2\times2=4$ atoms) was set up for three W mole fraction values of $x$, with $x=0025$, $0.5$, and $0.75$. Our choice of the $2\times2$ supercell is reasonable as the same trend is attained as obtained by the $9\times9$ supercell mentioned in a previous work~\cite{a06}. Figure~\ref{fig:1} shows a $2\times2$ supercell of monolayer Mo$_{0.75}$W$_{0.25}$S$_2$ in the plane adopted in this work.

\begin{figure}[t]
\includegraphics[width=3.3in]{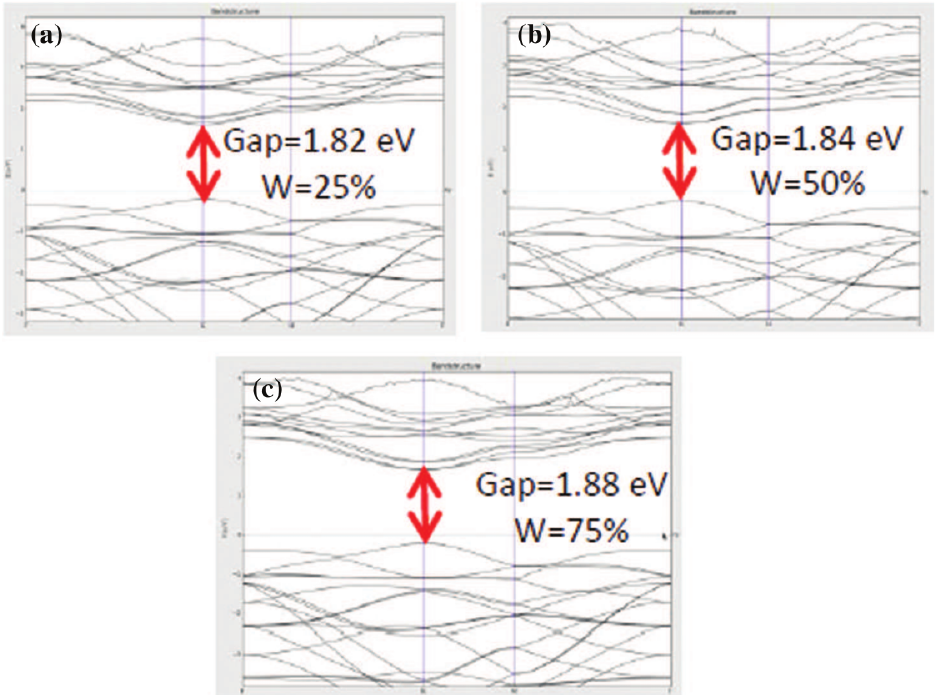}
\caption{Band structures of single layer Mo$_{1-x}$W$_x$S$_2$ at different W compositions. (a) W=25\%, band gap is $1.82$~eV, (b) W=50\%, band gap is $1.84$~eV, and (c) W=75\%, band gap is $1.88$~eV.}\label{fig:2}
\end{figure}

\begin{figure}[b]
\includegraphics[width=3.3in]{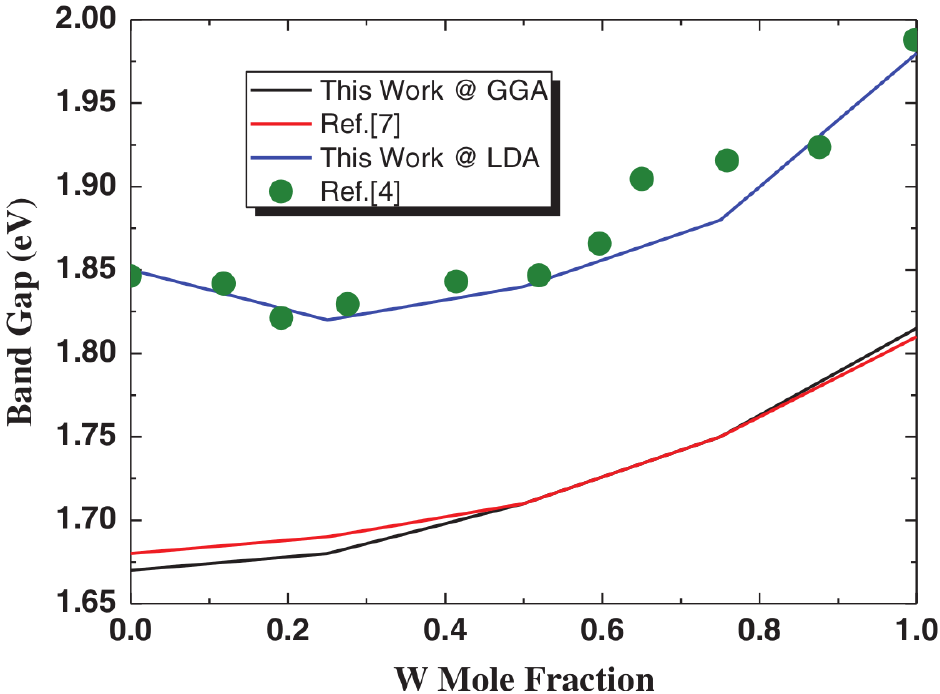}
\caption{Band gap versus W mole fraction for single layer Mo$_{1-x}$W$_x$S$_2$ alloys. Green dots are experimental results from a previous work~\cite{a04}.}\label{fig:3}
\end{figure}

In Figure~\ref{fig:2}, there are three first principle band structures of single layer alloy Mo$_{1-x}$W$_x$S$_2$ with three W values of 25, 50, and 75\%. The band gap of single layer MoS$_2$ could be varied by alloying transition metal atoms. The Mo$_{1-x}$W$_x$S$_2$ solutions remain as the direct bandgap semiconductors and the constituents of monolayer WS$_2$ and monolayer MoS$_2$. Band gap of single layer alloy Mo$_{1-x}$W$_x$S$_2$ versus W composition is shown in Figure~\ref{fig:3}. Note that with increasing W mole fraction, initially, the band gap of single layer alloy Mo$_{1-x}$W$_x$S$_2$ decreased gradually, and then from $x\approx0.25$ onward, a rapid increase was observed until the band gap reached that of the single layer WS$_2$. These results are consistent with the band gaps experimentally measured in terms of PL~\cite{a04}. Note that the generalized gradient approximation (GGA) considering the PBE function is typically adopted for the first-principles DFT calculation of the Mo$_{1-x}$W$_x$S$_2$ band gaps and for the comparison of results. A comparison indicated that our results are consistent with those of Wei et al.~\cite{a07}. We decided to use LDA to replace GGA in our work because LDA fitted experimental data better than did GGA.

\begin{figure}[t]
\includegraphics[width=3.3in]{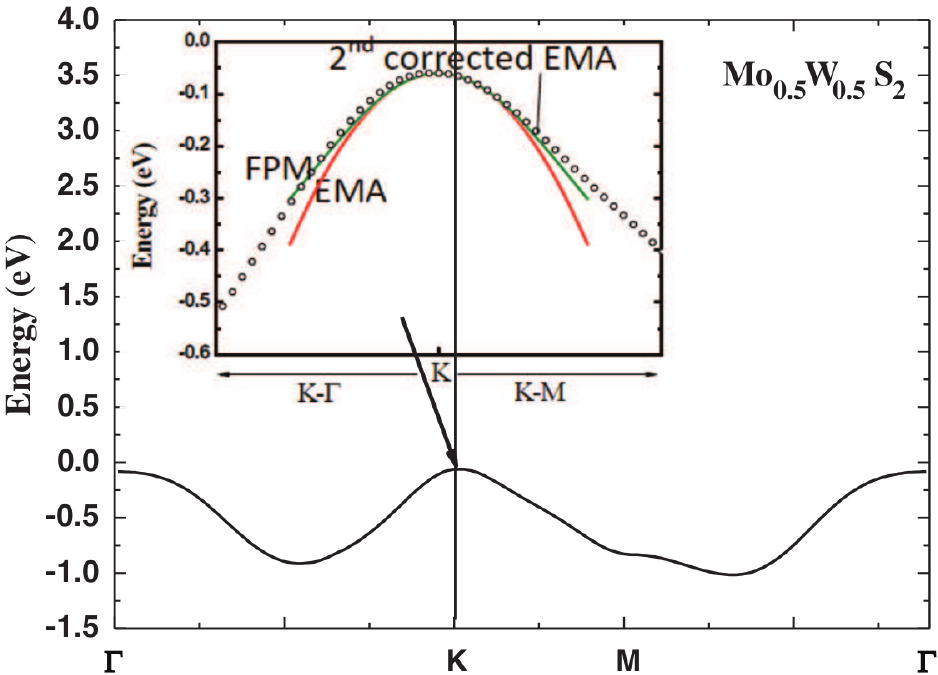}
\caption{Top valence band structure for single layer Mo$_{0.5}$W$_{0.5}$S$_2$ alloy from first-principles method (FPM) and conventional EMA model and EMA valence band model with the high-order nonparabolic correction.}\label{fig:4}
\end{figure}

\subsection{Compact Valence Bond Model}\label{sec:2-2}
As described in the previous subsection, the first principle band structure of single layer alloy Mo$_{1-x}$W$_x$S$_2$ was calculated considering the $2\times2$ supercell in calculation~\cite{a06}. Figure~\ref{fig:4} indicates that within an extremely small energy region, the valence band structure is ideal parabolic-like in $K$ valley due to the isotropic and parabolic dispersion feature of the holes in the valance band. The hole mobility is determined by the valence band described by a simple band model such as an effective mass approximation (EMA) band model with different effective mass along $x$ and $y$ direction, as shown in Eq.~\eqref{eq:1}.
\begin{equation}\label{eq:1}
    E=\frac{\hbar^2k_x^2}{2m_x^*}+\frac{\hbar^2k_y^2}{2m_y^*}
\end{equation}
where $k=(k_x,k_y)$ is the in-plane wave vector; $m_x^*$ and $m_y^*$ are, respectively, the effective masses in the top valence band including key $k$-points such as $K$, $\Gamma$ and $M$.

However, the simple EMA model cannot accurately determine the band structure of single layer alloy Mo$_{1-x}$W$_x$S$_2$ in the entire $k$-region. Therefore, it is required that the hole mobility of single layer Mo$_{1-x}$W$_x$S$_2$ be calculated using an improved compact valence band model. Hosseini et al.~\cite{a08} used simple band model considering the first-order non-parabolic factor $\alpha$, as shown in Eq.~\eqref{eq:2}:
\begin{equation}\label{eq:2}
    E\times(1+\alpha E)=\frac{\hbar^2}{2}\left(\frac{k_x^2}{m_x^*}+\frac{k_y^2}{m_y^*}\right)
\end{equation}

In our study, we adopted a higher-order correction for the simple band model of the hole band of silicon, as described in Ref.~\cite{a09}. Thus, we built a compact simple valence band model for single layer Mo$_{1-x}$W$_x$S$_2$ alloys by considering high-order non-parabolic factors $\alpha$ and $\beta$,as shown in Eq.~\eqref{eq:3}:
\begin{equation}\label{eq:3}
    E\times(1+\alpha E+\beta E^2)=\frac{\hbar^2}{2}\left(\frac{k_x^2}{m_x^*}+\frac{k_y^2}{m_y^*}\right)
\end{equation}
$\alpha$ and $\beta$ were obtained by fitting the first principle band structure for the top valence band.

\begin{figure}[t]
\includegraphics[width=3.3in]{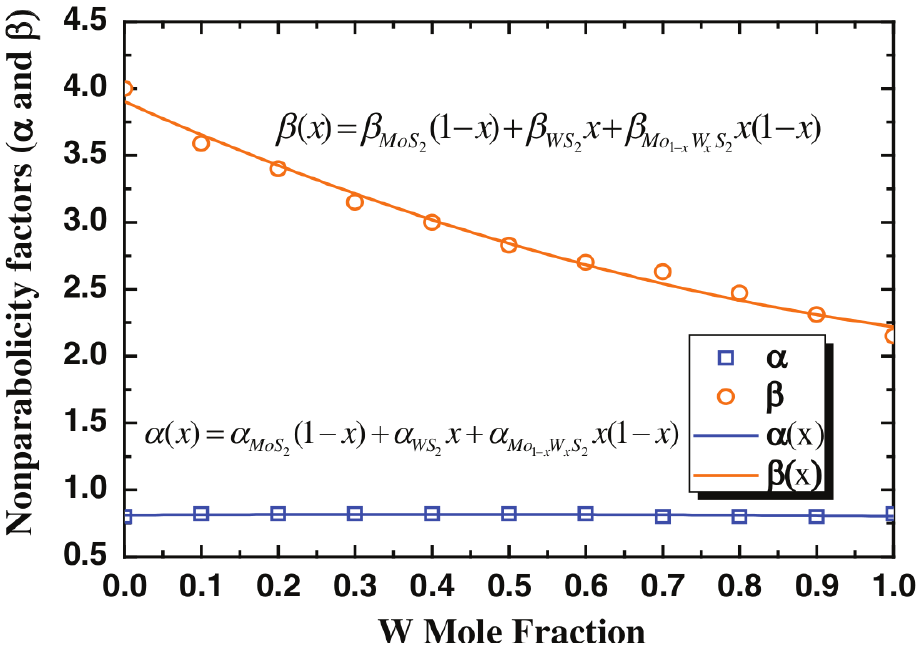}
\caption{Two nonparabolic factors ($\alpha$ and $\beta$) for holes in the EMA model with the high-order nonparabolic correction as a function of W mole fraction. Analytical functions for fitting the data, $\alpha(x)$ and $\beta(x)$, are also included for comparison.}\label{fig:5}
\end{figure}

\begin{figure}[b]
\includegraphics[width=3.3in]{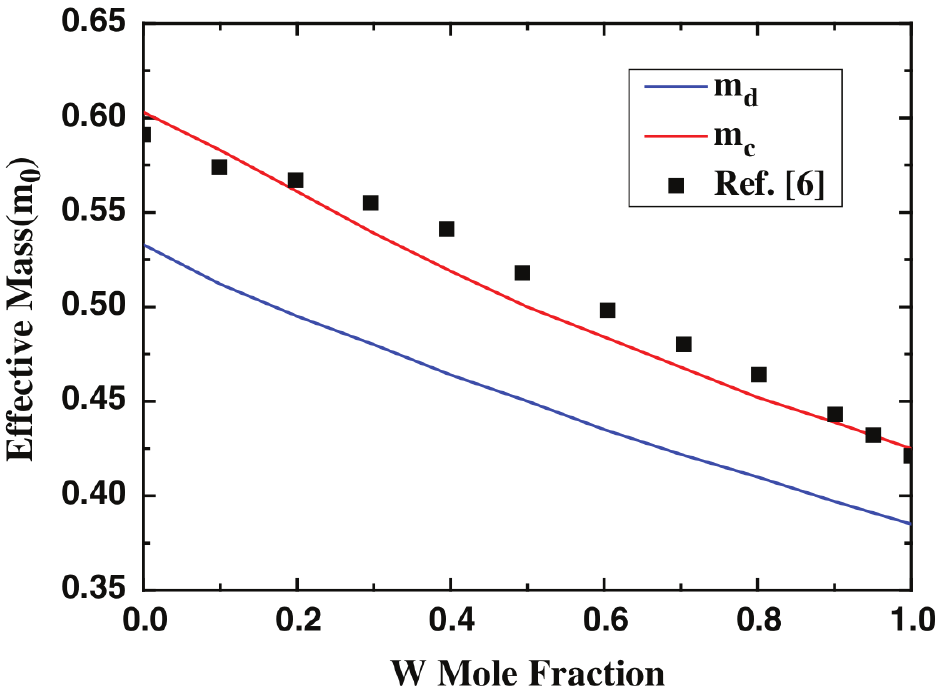}
\caption{Two types of effective masses including $m_C$ and $m_d$ as a function of W mole fraction. Note that $m_c$ is conductivity mass and $m_d$ is density of states mass, respectively. Calculated results from Ref.~\cite{a06} are also included for comparison. This work is in agreement with Ref.~\cite{a06}.}\label{fig:6}
\end{figure}

\subsection{Hole Mobility Calculation for Single Layer Alloy $\mathbf{Mo}_{1-x}\mathbf{W}_x\mathbf{S}_2$}\label{sec:2-3}
Based on an established method of semiconductor materials carrier mobility calculation~\cite{a10}, we calculated the hole mobility of Mo$_{1-x}$W$_x$S$_2$ alloy using the Kubo-Greenwood formula, given by Eqs.~\eqref{eq:4}-\eqref{eq:6}.
\begin{equation}\label{eq:4}
    \mu_{xx}=\frac{e\int_{E_0}^\infty D_{\mathrm{MoWS}_2}(E)V_x^2(E)\tau_\mathrm{tot}(E)f_\mathrm{FD}(E)[1-f(E)]dE}
    {k_BT\int_{E_0}^\infty D_{\mathrm{MoWS}_2}(E)f_\mathrm{FD}(E)dE}
\end{equation}
\begin{equation}\label{eq:5}
    D_{\mathrm{MoWS}_2}(E)=\frac{1}{(2\pi)^2}\int_{BZ}d^2\vec k\delta(E-E(\vec k))
\end{equation}
\begin{equation}\label{eq:6}
    V_x^2(E)=\frac{\frac{1}{(2\pi)^2}\int_{BZ}d^2\vec kV_x^2(\vec k)\delta(E-E(\vec k))}
    {\frac{1}{(2\pi)^2}\int_{BZ}d^2\vec k\delta(E-E(\vec k))}
\end{equation}
where $D_{\mathrm{MoWS}_2}(E)$ denotes the density of states for single layer alloy Mo$_{1-x}$W$_x$S$_2$, $V_x(E)$ is the group velocity with energy dependence along the $x$ direction~\cite{a11}, $f_\mathrm{FD}(E)$ means the Fermi–Dirac distribution, $E$ denotes carrier energy, and $\tau_\mathrm{tot}(E)$ as shown in Eq.~\eqref{eq:4} means the total scattering time. Further, $V(\vec k)=(\nabla_{\vec k}E(\vec k))/\hbar$ denotes $k$ dependent group velocity and $x$ means carrier transport direction. The different scattering rates from phonon and alloy effects are summed using Matthiessen's rule in order to determine the total scattering rate and the impact of various scattering mechanisms including phonon and alloy scattering:

\begin{equation}\label{eq:7}
    \frac{1}{\tau_\mathrm{tot}(E)}=\frac{1}{\tau_\mathrm{adp}(E)}
    +\frac{1}{\tau_\mathrm{odp}(E)}+\frac{1}{\tau_\mathrm{alloy}(E)}
\end{equation}
\begin{figure}[t]
\includegraphics[width=3.3in]{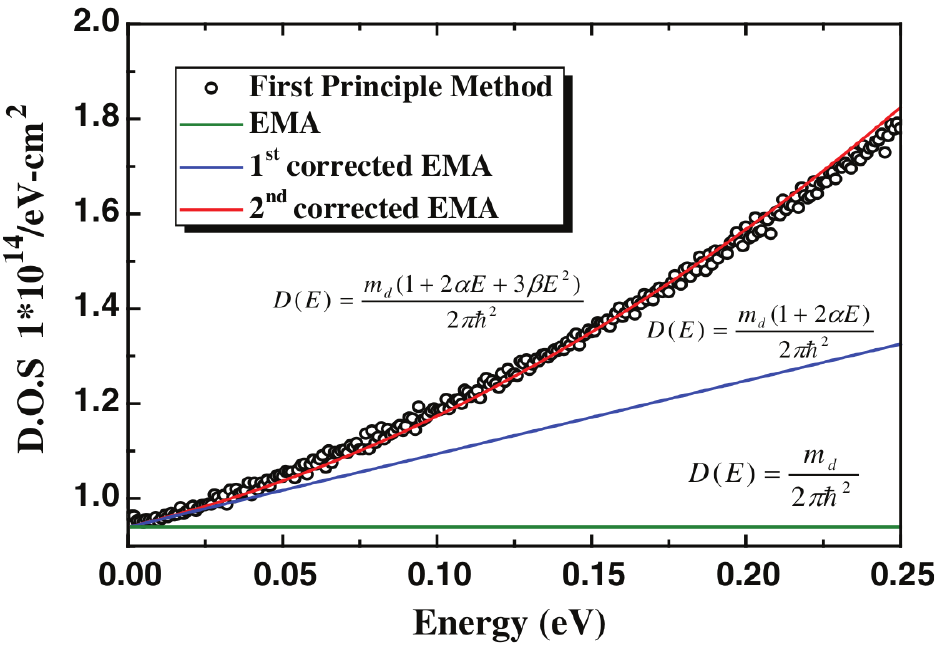}
\caption{Density of states as a function of carrier energy for single layer Mo$_{0.5}$W$_{0.5}$S$_2$. Open circles are first-principles results. Lines are results from three EMA models. Analytical formulas for density of states from three EMA models are also included in the inset.}\label{fig:7}
\end{figure}
\begin{figure}[b]
\includegraphics[width=3.3in]{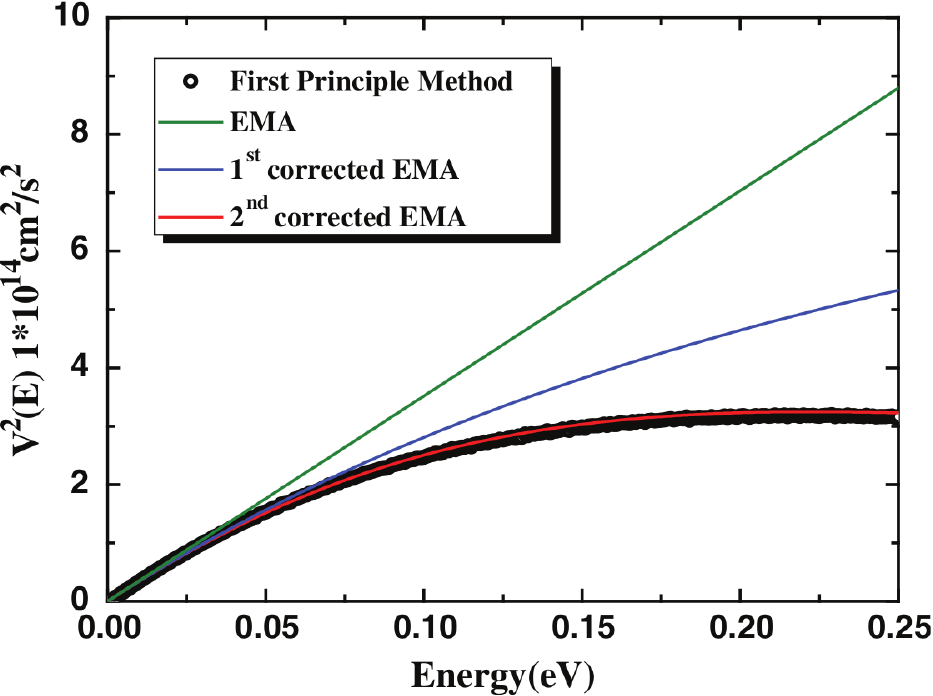}
\caption{Velocity square as a function of carrier energy for single layer Mo$_{0.5}$W$_{0.5}$S$_2$. Open circles are first-principles results. Lines are results from three EMA models. Analytical formulas for density of states from three EMA models are also included in the inset.}\label{fig:8}
\end{figure}

Here, $\frac{1}{\tau_\mathrm{alloy}(E)}$ is owing to alloy scattering, whereas the remaining two terms are related to phonon scattering. Moreover, the scattering rates owing to optical and acoustic phonon scattering, namely $\frac{1}{\tau_\mathrm{odp}(E)}$ and $\frac{1}{\tau_\mathrm{adp}(E)}$, respectively, for single layer Mo$_{1-x}$W$_x$S$_2$ alloys were adopted from the previous study~\cite{a12}. We followed the alloy scattering formula proposed by Ref.~\cite{a13} to calculate alloy scattering rate, and scattering formula is given by Eq.~\eqref{eq:8}.
\begin{equation}\label{eq:8}
    \frac{1}{\tau_\mathrm{alloy}(E)}=\frac{2\pi}{\hbar}\Delta U^2_\mathrm{alloy}x(1-x)A_{\mathrm{Mo}_{1-x}\mathrm{W}_x\mathrm{S}_2} D_{\mathrm{Mo}_{1-x}\mathrm{W}_x\mathrm{S}_2}(E)
\end{equation}
where $\Delta U^2_\mathrm{alloy}$ is the alloy scattering potential, which can be determined by the difference in the valence band offsets between monolayer MoS$_2$ and WS$_2$ or the metal work function difference between Mo and W. $A_{\mathrm{Mo}_{1-x}\mathrm{W}_x\mathrm{S}_2}$ denotes the area of the unit cell and $x$ means alloy composition of W in single layer alloy Mo$_{1-x}$W$_x$S$_2$.

\section{Results and Discussion}\label{sec:3}
The top valence band of single layer alloy Mo$_{0.5}$W$_{0.5}$S$_2$ calculated using the $2\times2$ supercell first-principle method is shown in Figure~\ref{fig:4}. Further, we compare the results of the conventional simple band model and the proposed improved model with the high-order non-parabolic effect. There is no major difference in the velocity square and density of states calculations between the two methods, thus indicating that the proposed compact band model with the high-order non-parabolic effect provides effective and better mobility calculation without comprising accuracy. Figure~\ref{fig:5} shows the plot of the non-parabolic factors $\alpha$ and $\beta$ calculated by Eq.~\eqref{eq:3} with respect to the mole fraction of W in single layer Mo$_{1-x}$W$_x$S$_2$. As you can see that $\beta$ is dependent on alloy composition of W and $\alpha$ is nearly independent of W. The inset of Figure~\ref{fig:5} shows the data fitted using a high-order polynomial. Note that no severe bowing effect is observed, especially for the non-parabolic factor $\alpha$.

\begin{figure}[t]
\includegraphics[width=3.3in]{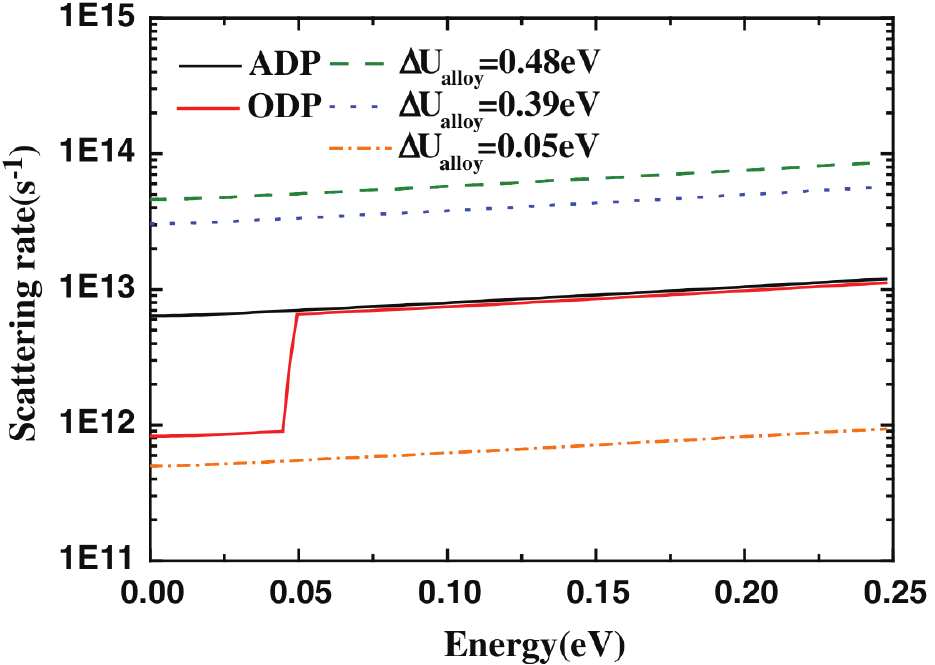}
\caption{Comparison of scattering rates from various phonon scattering mechanisms at a temperature of 300~K. Note that the scattering rate is calculated from first-principles valence band structure.}\label{fig:9}
\end{figure}

\begin{figure}[t]
\includegraphics[width=3.3in]{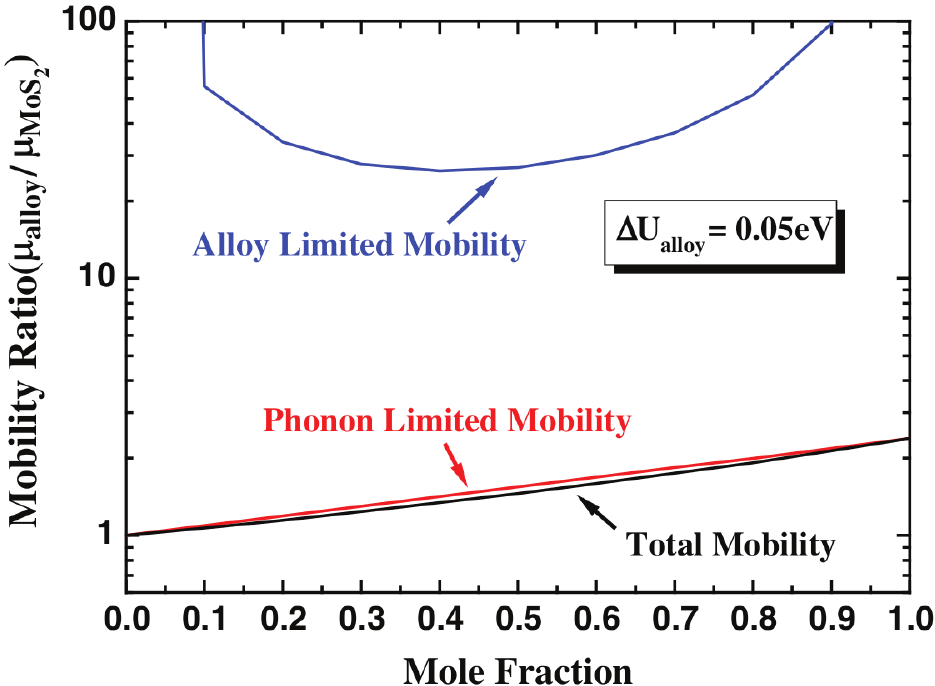}
\caption{Hole mobility contributed from alloy-limited, phononlimited, and total scattering as a function of W mole fraction from the compact valence band model of EMA with high-order nonparabolic correction with $\Delta U_\mathrm{alloy}=0.05$~eV. Note that the carrier concentration is fixed at $10^{12}$ cm$^{-2}$.}\label{fig:10}
\end{figure}

Figure~\ref{fig:6} shows the variation of the hole effective mass, namely conductivity effective mass (denoted as $m_c$; Eq.~\eqref{eq:3}, where $m_c$ is $m_x^*$)and density of states effective
mass (denoted as $m_d$; Eq.~\eqref{eq:3}, $md=\sqrt{m_x^*m_y^*}$), versus W mole fraction. The figure indicates that both effective mass values decrease almost linearly with increasing alloy composition of W owing to higher effective mass of single layer MoS$_2$ as compare to that of single layer WS$_2$. In addition, $m_c$ is higher than $m_d$ possibly because of the trigonal warping effect of the valence band.

Figures~\ref{fig:7} and \ref{fig:8} present velocity square and density of states values versus energy for monolayer alloy Mo$_{0.5}$W$_{0.5}$S$_2$, respectively. The figures indicate that velocity square and density of states calculations using the simple band model with the high-order non-parabolic effect agree well with the results from the first-principles method.
Therefore, the simple EMA band model with higher order non-parabolic effect is the optimal option to calculate hole mobility.

Futher, Figure~\ref{fig:9} shows the comparison of two hole scattering rates such as optical and acoustic phonon scattering and alloy scattering with three different alloy potentials.

\begin{figure}[t]
\includegraphics[width=3.3in]{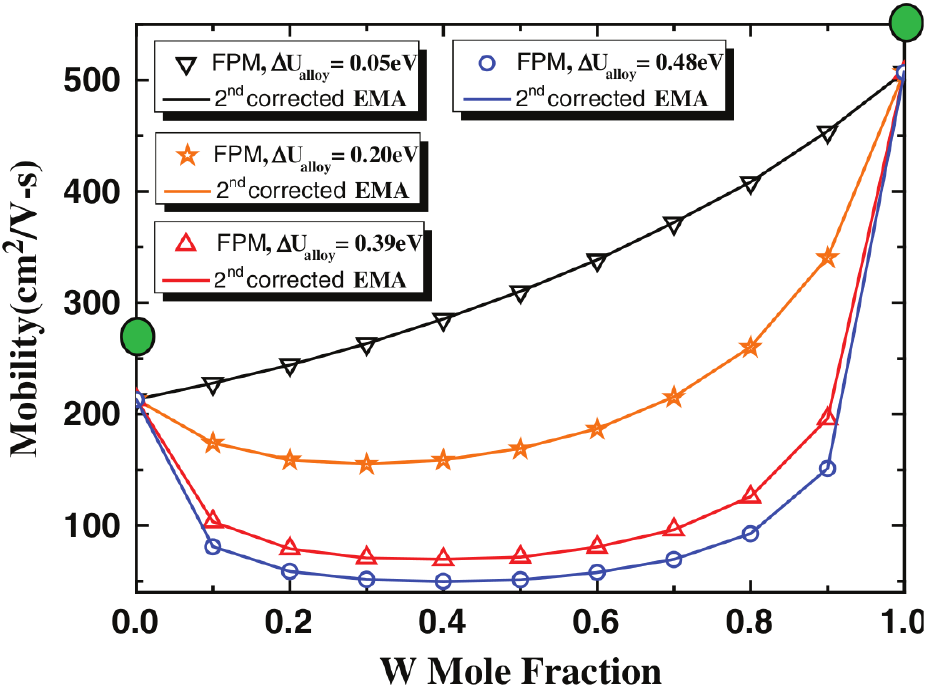}
\caption{Total mobility ratio from first-principles method (FPM) and the compact valence band model of EMA with high-order nonparabolic correction as a function of W mole fraction with four different $\Delta U_\mathrm{alloy}$ values. Theoretical results from Ref.~\cite{a14} (green circle) for MoS$_2$ and WS$_2$ are also included for comparison.}\label{fig:11}
\end{figure}

Figure~\ref{fig:10} shows alloy limited mobility, phonon limited mobility and total mobility ratios of the single layer alloy Mo$_{1-x}$W$_x$S$_2$ with $\Delta U_\mathrm{alloy}=0.05$~eV. We found that phonon limited mobility ratio increases up to $\approx2.5$ when W mole function is 100\%, while it was less than the alloy limited mobility ratio for the entire range of W mole fractions. Thus, the effect of alloy scattering is not prominent owing to phonon scattering is higher than alloy scattering. The hole mobility versus alloy composition of W considering different values of alloy scattering potential is shown in Figure~\ref{fig:11}. As you can see in Figure~\ref{fig:11}, alloy scattering potential values of $0.39$ and $0.48$~eV were adopted from the valence band offsets between MoS$_2$ and WS$_2$~\cite{a15,a16}. We also compare our results with the theoretical results of Ref.~\cite{a14}. Further, our results agree with those of Ref.~\cite{a14}. From Figure~\ref{fig:11}, we observe that the hole mobility for the single layer Mo$_{1-x}$W$_x$S$_2$ with $\Delta U_\mathrm{alloy}=0.05$~eV is the best because of the negligible alloy-limited scattering. A higher $\Delta U_\mathrm{alloy}$ value of $0.48$~eV results in mobility less than~1 (with respect to MoS$_2$), mainly because of the severe alloy scattering and lower hole mobility than in case of pure single layer MoS$_2$.

\section{Conclusions}\label{sec:4}
We investigated the alloy effect on the hole mobility of single layer alloy Mo$_{1-x}$W$_x$S$_2$. We determined the band gap of single layer alloy Mo$_{1-x}$W$_x$S$_2$ from $2\times2$ supercell first-principle calculation; further, we used a simple band model as well as a hole mobility calculation for single layer alloy Mo$_{1-x}$W$_x$S$_2$. The simple valence band model can be applied to the TCAD design study of next generation electron devices that use monolayer molybdenum tungsten alloy disulfide as the channel material.

\begin{acknowledgements}
Authors all appreciate supporting of the Ministry of Science and Technology, Taiwan, R.O.C., under contract numbers MOST 108-2622-8-002-016, 106-2221-E-005-096-MY3, and 106-2221-E-005-097-MY3 and also thank to Uni-edit (\href{http://www.uni-edit.net}{www.uni-edit.net}) for English proofreading and editing this manuscript.
\end{acknowledgements}

%
\end{document}